\documentclass[nojss]{jss}

\usepackage{url}		
\usepackage{xspace}		
\usepackage{booktabs}
\usepackage{amsmath}
\usepackage{amsfonts}
\usepackage{rotating}	
\usepackage{pdflscape}
\usepackage{fancyvrb}

\usepackage{bbm}			

\newtheorem{prop}{Proposition}

\newcommand{\cADBH}{\code{ADBH}\xspace}
\newcommand{\cDBH}{\code{DBH}\xspace}
\newcommand{\cDBR}{\code{DBR}\xspace}
\newcommand{\cDiscBH}{\code{discrete.BH}\xspace}

\newcommand{\nat}{{\mathbb N}}

\newcommand{\PoisVert}{\textbf{{Pois}}}

\newcommand{\FDR}{\textnormal{FDR}}

\newcommand{\xiSD}{\xi_{\textnormal{SD}}}
\newcommand{\xiSU}{\xi_{\textnormal{SU}}}

\defcitealias{DDR2018}{[DDR]}
\defcitealias{Dickhaus2012}{Dickhaus, Stra{\ss}burger et al. (2012)}
\defcitealias{Karp2016}{Karp, Heller et al. (2016)}
\defcitealias{vandenBroek2015}{van den Broek, Dijkstra et al. (2015)}
\defcitealias{RMutoss2017}{Blanchard, Dickhaus et al. (2017)}

\usepackage{thumbpdf, lmodern}

\usepackage{framed} 
\usepackage{longtable} 
\usepackage{multirow}
\usepackage{float}
\usepackage{appendix}
\usepackage{graphicx}

\author{Guillermo Durand\\Sorbonne University
   \And 
   Florian Junge\\Darmstadt University\\of Applied Sciences
   \And
   Sebastian D\"ohler\\Darmstadt University\\of Applied Sciences
   \And Etienne Roquain\\Sorbonne University}
\Plainauthor{Guillermo Durand, Florian Junge, Sebastian D\"ohler,  Etienne Roquain}

\title{\pkg{DiscreteFDR}: An \proglang{R} package for controlling the false discovery rate for discrete test statistics}
\Plaintitle{DiscreteFDR R package}
\Shorttitle{DiscreteFDR \proglang{R} package}

\Abstract{
The simultaneous analysis of many statistical tests is ubiquitous in applications. Perhaps the most popular error rate used for avoiding type one error inflation is the false discovery rate (FDR). However, most theoretical and software development for FDR control has focused on the case of continuous test statistics. For discrete data, methods that provide proven FDR control and good performance have been proposed only recently. The \proglang{R} package \pkg{DiscreteFDR} (\cite{DJDR2018}, version 1.2) provides an implementation of these methods. For particular commonly used discrete tests such as Fisher's exact test, it can be applied as an off-the-shelf tool by taking only the raw data as input. It can also be used for any arbitrary discrete test statistics by using some additional information on the distribution of these statistics. The paper reviews the statistical  methods in a non-technical way, provides a detailed description of the implementation in \pkg{DiscreteFDR} and presents some sample code and analyses. 
}

\Plainkeywords{FDR, package, R}

\Address{
  Guillermo Durand\\
  Sorbonne Universit\'e,\\
  Laboratoire de Probabilit\'es, Statistique et Mod\'elisation\\
  4, place Jussieu\\
  75005 Paris, France\\
  E-mail: \email{Guillermo.Durand@upmc.fr}\\
  
  Florian Junge\\
  Darmstadt University of Applied Sciences\\
  Darmstadt Institute for Statistics and Operations Research\\
  Sch\"offerstra{\ss}e 3\\
  64295 Darmstadt, Germany\\
  E-mail: \email{florian.junge@h-da.de}\\
  
  Sebastian D\"ohler\\
  Darmstadt University of Applied Sciences\\
  Faculty of Mathematics and Natural Sciences\\
  Darmstadt Institute for Statistics and Operations Research\\
  Sch\"offerstra{\ss}e 3\\
  64295 Darmstadt, Germany\\
  E-mail: \email{sebastian.doehler@h-da.de}\\
  
  Etienne Roquain\\
  Sorbonne Universit\'e,\\
  Laboratoire de Probabilit\'es, Statistique et Mod\'elisation\\
  4, place Jussieu\\
  75005 Paris, France\\
  E-mail: \email{Etienne.Roquain@upmc.fr}\\
}

\begin{document}





\section[Introduction]{Introduction} \label{sec:intro}

Multiple testing procedures are important tools for identifying statistically significant findings in massive and complex data while controlling a specific error rate. An important focus has been given to methods controlling the false discovery rate (FDR), i.e., the expected proportion of falsely rejected hypotheses among all rejected hypotheses, which has become the standard error rate for high dimensional data analysis. Since the original procedure of \cite{BenjaminiHochberg95}, much effort has been undertaken  to design FDR controlling procedures that adapt to various underlying structures of the data, such as the quantity of signal, the signal strength and the dependencies, among others. 

\clearpage

The \proglang{R} package \pkg{DiscreteFDR}, presented in this paper, deals with adaptation  to discrete and non-identically distributed test statics by implementing procedures developed by \cite{DDR2018} (in the sequel abbreviated as \citetalias{DDR2018}). This type of data arises in many relevant applications, in particular when data represent frequencies or counts. Examples can be found in  clinical studies (see e.g., \cite{WestWolf1997}), genome-wide association studies (GWAS)  (see e.g., \citetalias{Dickhaus2012}) and next generation sequencing data (NGS)   (see e.g., \cite{DoergeChen2015}). The primary discrete test we have in mind in this paper is Fisher's exact test, see \cite{LehmannRomano}, but we also sketch an application of \pkg{DiscreteFDR} to multiple Poisson tests in the spirit of \cite{JimenezUnaAlvarez2018}.

It is well known (see e.g., \cite{WestWolf1997}) that applying critical values derived for continuous approximations to discrete null distributions can generate a severe power loss, already at the stage of the single tests. A consequence is that using 'blindly' the BH procedure with discrete $p$-values will control the FDR in a  too conservative manner. Therefore, more powerful procedures that avoid this conservatism are much sought after  in applications, see for instance \citetalias{Karp2016}, \citetalias{vandenBroek2015} and \citetalias{Dickhaus2012}.

\begin{figure}[htbp]
	\centering
	\vspace{-1.5cm}
	\includegraphics[width=1\textwidth]{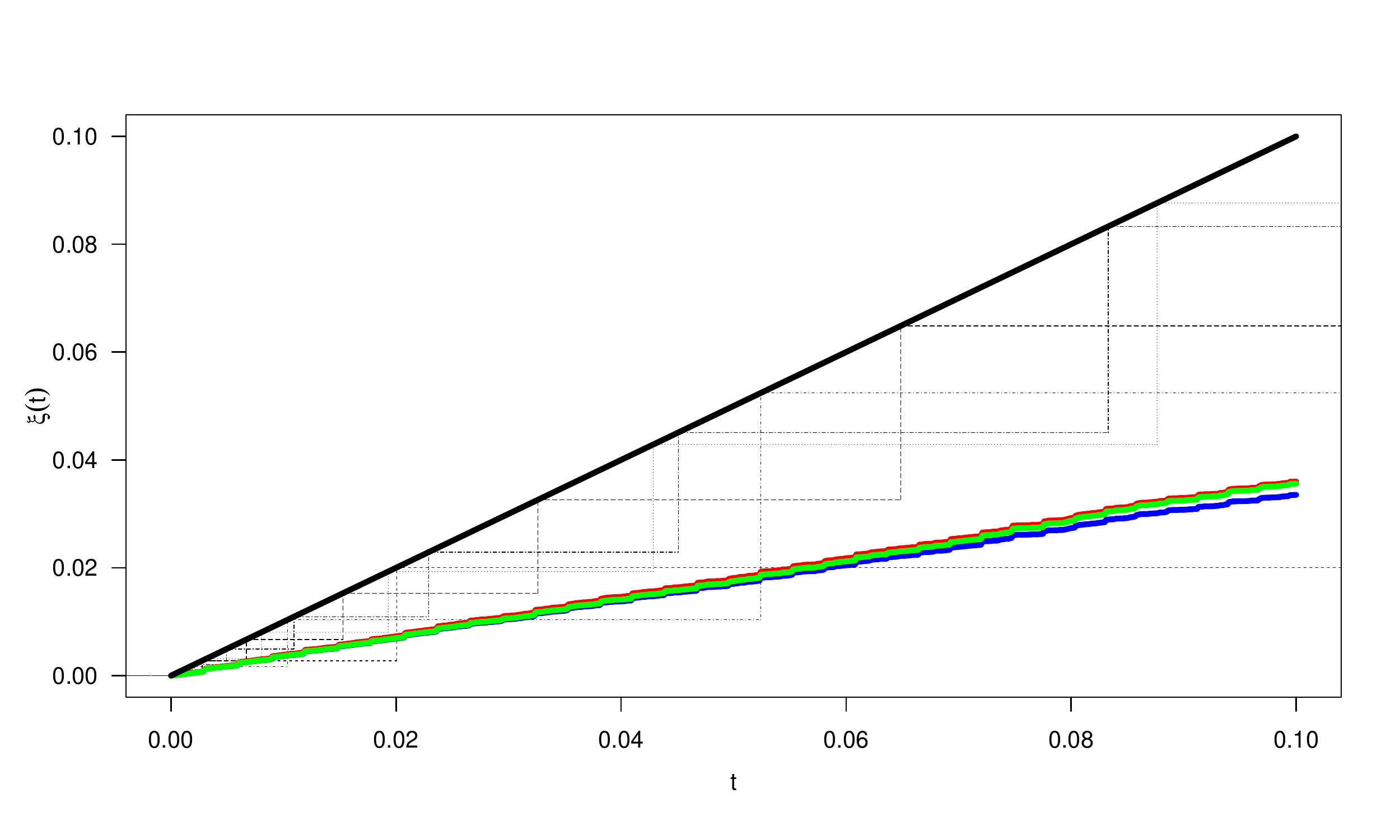}
	\vspace{-1.5cm}
	\caption{Plots of variants of $\xi$ for the pharmacovigilance data. The solid black line corresponds to the uniform case, the discrete variants are represented by blue (for $\xi$), green (for $\xiSD$) and  red (for $\xiSU$) solid lines. Additionally, five arbitrarily selected $F_i$'s are displayed by using different line types.}
	\label{fig:FbarPlotsHellerForPaper}
\end{figure}

In the literature, constructing multiple testing procedures that take into account the discreteness of the test statistics has a long history, for more details see \citetalias{DDR2018}. The heuristic motivation for the procedures implemented in \pkg{DiscreteFDR} is as follows. Let $p_{(1)} \le \ldots \le p_{(m)} $ denote the ordered $p$-values and $H_{(1)},\ldots,H_{(m)} $ the corresponding null hypotheses. The BH procedure [BH] works by rejecting $H_{(1)},\ldots,H_{(\hat{k})} $, where $\hat{k}$ is the largest integer $k$ such that 
\begin{align}
p_{(k)} & \le \frac{k}{m}\cdot \alpha. \label{eq:BH:procedure}
\end{align}
Now suppose that the cumulative distribution functions $F_1, \ldots,F_m$ of the $p$-values under the null hypotheses are known and introduce the transformation 
\begin{align}
\xi(t) & = \frac{1}{m} \sum_{i=1}^m F_i(t) , \:\:t\in[0,1]. \label{eq:Heyse0}
\end{align}
For  continuous settings we often have $F_i(t)=t$ which implies $\xi(t)=t$ and so we can rephrase \eqref{eq:BH:procedure} as 
\begin{align}
\xi(p_{(k)}) & \le \frac{k}{m}\cdot \alpha. \label{eq:Heyse}
\end{align}
\cite{Heyse2011} proposed to use the transformation $\xi$ in \eqref{eq:Heyse0}, where the $F_i$ need no longer be uniform and identical. The benefit of this approach is that - depending on the discreteness and heterogeneity of the involved $p$-value distributions - $\xi(t)$ may be much smaller than $t$. Clearly, the smaller the $\xi$-values, the more hypotheses can be rejected. Figure~\ref{fig:FbarPlotsHellerForPaper} displays such a  function where the functions $F_1, \ldots, F_{2446}$ are derived from $m=2446$ independent Fisher's exact test statistics based on the pharmacovigilance data from \citet{Heller2012} (see Section~\ref{sec:Using}  for more details). In this example we have $\xi(t) \approx t/3$, thereby  yielding a potentially strong rejection enhancement. 

Unfortunately, the Heyse procedure does not rigorously control the FDR in general; counter-examples are provided in  \cite{Heller2012} and \citetalias{DDR2018}. To correct this, \citetalias{DDR2018} introduce new procedures relying on the following modifications  of the $\xi$ function (more details are presented in Section \ref{sec:mathematics}):
\begin{align*}
	\xiSU (t)= \frac{1}{m}\sum_{i=1}^m \frac{F_i\left(t\right)}{1-F_i\left(\tau_m\right)};\:\:\:
	\xiSD (t)=\frac{1}{m}\sum_{i=1}^m \frac{F_i\left(t\right)}{1-F_i\left(t\right)},\:\: t\in[0,1],
\end{align*}
where $\tau_m$ is the generalized inverse of $\xiSD$ taken at point $\alpha$. Figure~\ref{fig:FbarPlotsHellerForPaper} demonstrates that the difference between these modifcations and the original $\xi$ can be very small, in particular for small values of $t$. In addition, \citetalias{DDR2018} also introduce more powerful 'adaptive' versions, meaning that the derived critical values are designed in a way that 'implicitly estimates' the overall proportion of true null hypotheses. All these procedures provide rigorous $\FDR$ control under independence of the $p$-values and are implemented in the \proglang{R} package \pkg{DiscreteFDR}.

While there exist numerous \proglang{R} functions and packages  that implement multiple testing procedures in the continuous setting (see e.g., \cite{RMultcomp2008} and \citetalias{RMutoss2017}), there are only relatively few tools available that deal specifically with discrete test statistics. The package \pkg{MHTdiscrete} (see \cite{MHTdiscrete}) is described by its authors as a 'comprehensive tool for almost all existing multiple testing methods for discrete data'. It implements several FWER and FDR procedures designed for discrete data. While the procedures for FWER control are extensively described in an accompanying preprint (see \cite{ZhuGuo2017}), there seems to be no detailed mathematical description of the implemented FDR procedures. The package \pkg{discreteMTP} (see \cite{discreteMTP}) also implements several methods aiming at $\FDR$ control (including the Heyse procedure) described in more detail in \cite{Heller2012}. The main contribution  of the package \pkg{DiscreteFDR} is to  provide practitioners with a simple to use set of tools (both adaptive and non-adaptive) for analysing discrete data with both proven $\FDR$ control and good performance.

In this paper, our primary aim is to introduce \pkg{DiscreteFDR}. As an 'appetizer', we start by illustrating the main ideas through analysis of a  toy data set. We hope to convince readers that it is worthwile to use discrete FDR methods. We then review the mathematical methods and results from \citetalias{DDR2018}, followed by  some more technical details of the implementation in   Section \ref{sec:implementation} . Section \ref{sec:Using} contains an analysis of some real data and includes an example that illustrates how \pkg{DiscreteFDR} can be used for arbitrary discrete tests. The paper concludes with a summary and discussion. 

We realize - and indeed hope - that the audience of this paper may be quite heterogeneous, which is why we would like to suggest some guidance for possible ways of reading it. For subject matter scientists and practitioners who may not be interested in the mathematical or software details, we especially recommend to study Sections \ref{sec:ToyExample} and \ref{sec:Using}. For readers who additionally want to understand more of the mathematical background we recommend Section \ref{sec:mathematics}, for readers interested in the implementation details of the R-package we recommend Section \ref{sec:implementation}.


\section[Examples]{A toy example} \label{sec:ToyExample}
To give a first impression of how  \pkg{DiscreteFDR} works, we consider an  artifical toy example. A more realistic example involving pharmacovigilance data is given in Section \ref{sec:Using}.

Suppose we would like to compare two treatments in nine different populations. For each population we do this by evaluating the responders and non-responders for each treatment. This leads to categorical data which can be  represented, for each population $i=1, \ldots,9$ in the following  2 $\times$ 2 table:

\begin{table}[htb]
\begin{center}
\begin{tabular}{lccc}
            & responders & non responders & total   \\
treatment $1$ & $x_{1i}$          & $y_{1i}$             & $n_{1i}$      \\
treatment $2$ & $x_{2i}$          & $y_{2i}$              & $n_{2i}$ \\
total       &  $x_{1i} + x_{2i}$   &  $y_{1i} + y_{2i}$   & $n_i = n_{1i} + n_{2i}$     
\end{tabular}
\caption{2 $\times$ 2 table for population $i$.}
\end{center}
\end{table}

Denoting the responder probabilities for population $i$ by $\pi_{1i}$ and $\pi_{2i}$ we can test e.g.
\begin{align*}
H_{0i}: \pi_{1i} = \pi_{2i} & \qquad \text{vs.} \qquad H_{1i}: \pi_{1i} \neq \pi_{2i}
\end{align*}
by using Fisher's (two-sided) exact test (see \cite{LehmannRomano}, which is implemented in the \proglang{R} function \code{fisher.test}). Suppose the data in the nine populations are independent and we observe the following data frame \code{df} 

\begin{figure}[htb]
\centering
\includegraphics{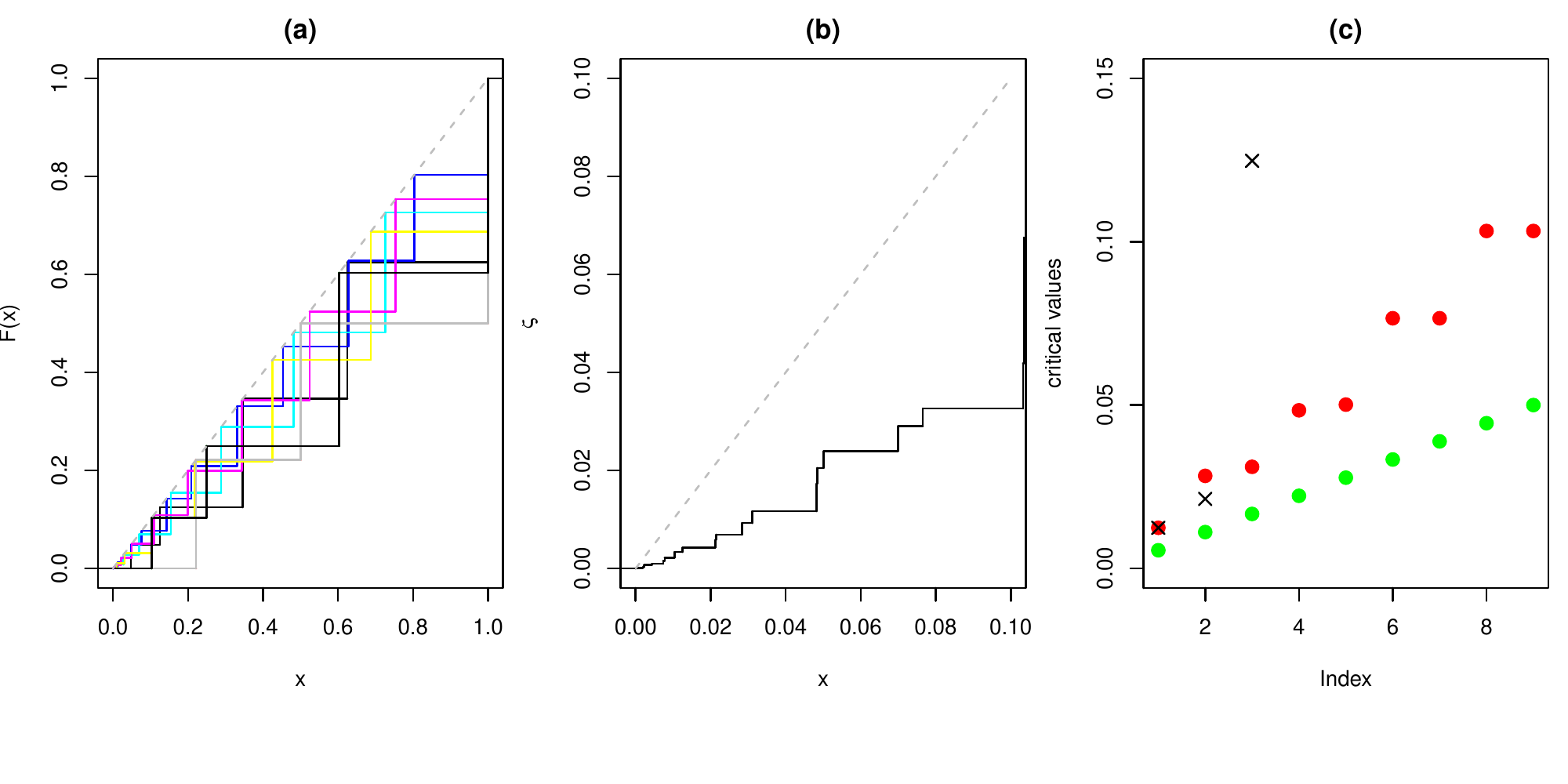}
\caption{\label{fig:otto} Panel (a) depicts the distribution functions $F_1, \ldots, F_9$ in various colours, (b) is a graph of the transformation $\xiSD$. The uniform distribution function is shown in light grey in (a) and (b). Panel (c) shows the [BH] critical values (green dots), the DBH step-down critical values (red dots) and the sorted raw $p$-values (asterisks).}
\end{figure}

\begin{Schunk}
\begin{Soutput}
  X1  Y1 X2  Y2
1  4 144  0 132
2  2 146  0 132
3  2 146  1 131
4 14 134  3 129
5  6 142  2 130
6  9 139  1 131
7  4 144  2 130
8  0 148  2 130
9  1 147  2 130
\end{Soutput}
\end{Schunk}
In this data frame each of the 9 rows  represents the data of an observed 2 $\times$ 2 table:  e.g., the third row of the data corresponds to $x_{13} = 2, y_{13} = 146, x_{23} = 1, y_{23} = 131$. Even though in this example, the total number of tested hypotheses $m=9$ is very small, for illustrative purposes we deal with the multiplicity problem here by controlling $\FDR$ at level $\alpha = 5\%$. The DBH step-down procedure (to be explained in more detail in Section \ref{sec:mathematics}) can be applied directly to the data frame object \code{df} and yields the following adjusted $p$-values:

\begin{Schunk}
\begin{Sinput}
R> library("DiscreteFDR")
R> DBH.sd.fast <- fast.Discrete(df, alternative = "two.sided", 
+                               direction = "sd")
R> DBH.sd.fast$Adjusted
\end{Sinput}
\begin{Soutput}
[1] 0.25630985 1.00000000 1.00000000 0.03819796 0.51482782 0.03819796
[7] 1.00000000 0.47895996 1.00000000
\end{Soutput}
\end{Schunk}
Thus we can reject two hypotheses at $\FDR$-level $\alpha=5\%$. In order to compare this with the usual (continuous) BH procedure we have to determine the raw $p$-values first. This would be possible by applying the \code{fisher.test} function to all  nine 2 $\times$ 2 tables. Alternatively, we may use the more convenient function \code{fisher.pvalues.support} included in our package for accessing the raw $p$-values:
\begin{Schunk}
\begin{Sinput}
R> p <- fisher.pvalues.support(df, alternative = "two.sided")
R> raw.pvalues <- p$raw
R> p.adjust(raw.pvalues, method = "BH")
\end{Sinput}
\begin{Soutput}
[1] 0.37430072 0.74976959 1.00000000 0.09570921 0.51928737 0.09570921
[7] 0.77313633 0.49804147 0.77313633
\end{Soutput}
\end{Schunk}
Applying the continuous BH procedure from the \pkg{stats} package in the last line of code, we find that we can not reject any hypotheses at $\FDR$-level $\alpha=5\%$. As this example illustrates, the discrete approach can potentially yield a large increase in power. The gain depends  on the involved distribution functions and the raw $p$-values. To appreciate where this  comes from, it is instructive to consider the distribution functions $F_1, \ldots, F_{9}$ of the $p$-values under the null in more detail. Take for instance the first 2 $\times$ 2 table: 

\begin{table}[htb]
\begin{center}
\begin{tabular}{lccc}
            & responders & non responders & total   \\
treatment $1$ & $4$          & $144$             & $148$      \\
treatment $2$ & $0$          & $132$              & $132$ \\
total       &  $4$   &  $276$   & $280$     
\end{tabular}
\caption{2 $\times$ 2 table for population $1$.}
\end{center}
\end{table}
Fisher's exact test works by determining the probability of observing this (or a more 'extreme') table, given that the margins are fixed. So each $F_i$ is determined by the margins of table $i$. Since $x_{11}+x_{21}=4$, the only potentially observable tables are given by $x_{11}=0, \ldots,4$. For each one of these 5 values  a $p$-value can be determined using the hypergeometric distribution. Therefore, the $p$-value of any 2 $\times$ 2 table with margins given by the above table can take (at most) 5 distinct values, say $x_1, \ldots, x_5$. Combining these 5 values into a set, we obtain the \emph{support} $\mathcal{A}_1= \{x_1, \ldots, x_5 \}$ of distribution $F_1$. Now we can continue in this vein for the remaining 2 $\times$ 2 tables to obtain the supports $\mathcal{A}_1, \ldots, \mathcal{A}_9$ for the distributions functions $F_1, \ldots, F_{9}$. The supports can be accessed via the \code{$support} command, e.g.
\begin{Schunk}
\begin{Sinput}
R> p$support[c(1,5)]
\end{Sinput}
\begin{Soutput}
[[1]]
[1] 0.04820493 0.12476691 0.34598645 0.62477763 1.00000000

[[2]]
[1] 0.002173856 0.007733719 0.028324482 0.069964309 0.154043258
[6] 0.288492981 0.481808361 0.726262402 1.000000000
\end{Soutput}
\end{Schunk}
returns $\mathcal{A}_1$ and $\mathcal{A}_5$. Panel (a) in Figure \ref{fig:otto} shows a graph of the distribution functions $F_1, \ldots, F_{9}$. Each $F_i$ is a step-function with $F_i(0)=0$, the jumps occuring only on the support $\mathcal{A}_i$ and $F_i(x)=x$ only for $x \in \mathcal{A}_i$. In particular, all distributions are stochastically larger than the uniform distribution (i.e., $F_i(x) \le x$), but in a heterogeneous manner. This heterogeneity can be exploited e.g., by transforming the raw $p$-values from the exact Fisher's test using the function $\displaystyle \xiSD  (x) = \sum_{i=1}^9 \frac{F_i(x)}{1-F_i(x)}$ presented in the Introduction.  Panel (b) shows that $\xiSD$ is a step function. Its jumps occur on the joint support $\mathcal{A}= \mathcal{A}_1 \cup \ldots \cup \mathcal{A}_9$. Panel (b) also shows that $\displaystyle \xiSD  (x) \ll x$, at least for small values of $x$. It turns out  that the critical values of our new DBH step-down procedure  are essentially given by inverting $\xiSD$ at the critical values of the [BH] procedure $1 \cdot \alpha/9, 2 \cdot \alpha/9, \ldots, \alpha $, so that these values are considerably larger than the [BH] critical values (for more details see Section \ref{sec:mathematics}). This is illustrated in panel (c) along with the ordered $p$-values. In particular, all asterisks are located above the green [BH] dots, therefore this procedure can not reject any hypothesis. In contrast, the two smallest $p$-values are located below red DBH step-down dots, so that this procedure rejects two hypotheses as we had already seen earlier.


\section[New procedures]{Implemented FDR-controlling procedures} \label{sec:mathematics}





 The procedures used in the package are all based upon a comparison between the ordered $p$-values $p_{(k)}$, $1\leq k \leq m$, and a sequence of nondecreasing {\it critical values} $\tau_{k}$, $1\leq k \leq m$. Depending on how these two sequences intercept allow to define a rejection number $\hat k$ and thus a rejection set of null hypotheses $H_{(1)},\dots,H_{(\hat{k})}$.
  Classically, the {\it step-up procedure} with critical values $\tau_{k}$, $1\leq k \leq m$, corresponds to choose the last right crossing point
  $$
  \hat{k}_{SU}=\max\{k\::\: p_{(k)}\leq \tau_k \}.
  $$
 Hence, it goes backwards, starting from the largest $p$-value $p_{(m)}$, stopping the first time it finds $k_0$ such that $p_{(k_0)}\leq \tau_{k_0}$ and returning $\hat{k}_{SU}=k_0$. 
 By contrast, the {\it step-down procedure} with critical values $\tau_{k}$, $1\leq k \leq m$ uses the first left crossing point
 $$
 \hat{k}_{SD}=\max\{k\::\: \mbox{ for all } k'\leq k,\:p_{(k')}\leq \tau_{k'} \}.
 $$
Hence, it goes forward, starting from the smallest $p$-value $p_{(1)}$, stopping the first time it finds $k_0$ such that $p_{(k_0)}> \tau_{k_0}$ and returning $\hat{k}_{SD}=k_0-1$. 

Such multiple testing procedures are thus driven by a sequence of critical values and by a choice between the step-up or step-down version. In our package, the $5$ different possible choices are listed in Table~\ref{tab:ListProceduresTransformations}, with $3$ step-up procedures [DBH-SU], [A-DBH-SU], [DBR-$\lambda$] and $2$ step-down procedures [DBH-SD], [A-DBH-SD]. We easily check that [A-DBH-SU] (resp. [A-DBH-SD]) rejects always more null hypotheses than [DBH-SU] (resp. [DBH-SD]).  
Note that the names of the procedures are slightly different in the original paper \citetalias{DDR2018}. This is done to emphasize that our package is primarily devoted to the discrete case.

\subsection{Critical values} \label{subsec:CritConsts}

The specific shape of the critical values comes from the FDR upper-bounds derived in \citetalias{DDR2018}, which ensures that these procedures control the FDR at the nominal level $\alpha$ under independence of the $p$-values, see Theorem~1 and Corollary~1 in \citetalias{DDR2018}.
In Table~\ref{tab:ListProceduresTransformations}, each $F_i$ is defined as the (least favorable) cumulative distribution function of the $p$-value $p_i$ under the null hypothesis.
As in the example from Section \ref{sec:ToyExample}, $\mathcal{A} =\mathcal{A}_1 \cup \ldots \cup \mathcal{A}_m \subset [0,1]$ stands for the union of the supports of the marginal distributions of the $p$-values, $p_i$, $1\leq i \leq m$ which can be determined under the full hull hypothesis, i.e., when all null hypotheses are assumed to be true. While $\mathcal{A}=[0,1]$ in the case where the $F_i$'s are continuous functions, the primary setting we have in mind is a large number of simultaneous Fisher exact tests, so that $\mathcal{A} =\mathcal{A}_1 \cup \ldots \cup \mathcal{A}_m $ is finite but very large. See also Section \ref{sec:ToyExample} for some concrete examples.

Let us underline that obtaining such $\tau_k$ numerically might be time consuming because the overall support $\mathcal{A}$ can be large while testing whether each $t\in \mathcal{A}$ satisfies the required condition given by the second column in Table \ref{tab:ListProceduresTransformations} involves a complex combination of the $F_i$, $1\leq i \leq m$. In the package, we have implemented a shortcut that reduces the range of $t\in \mathcal{A}$ that has to be explored: it is based on the fact that if $F_i(t)\leq t$ for all $t$ and $i$ (super-uniformity), 
we have the following lower bounds $\tau^{\tiny \mbox{min}}_k$'s on the critical values $\tau_k$'s, see Lemmas~2, 3 and 4 in \citetalias{DDR2018}:

\begin{tabular}{ll}
\toprule \relax
 [DBH-SU] &  $\tau^{\tiny \mbox{min}}_k = \max\{t\in \mathcal{A}\::\: t\leq \alpha k/m (1+\alpha)^{-1}  \}$
 \\ \relax
 [DBH-SD] &  $\tau^{\tiny \mbox{min}}_k = \max\{t\in \mathcal{A}\::\: t\leq \alpha k/m (1+\alpha k/m)^{-1}   \}$\\  \relax
  [A-DBH-SU] &
   {$\begin{aligned}[t]
  			\tau^{\tiny \mbox{min}}_m &=\max\{t\in \mathcal{A}\::\: t\leq \alpha (1+\alpha)^{-1}  \}\\
  			\tau^{\tiny \mbox{min}}_k &= \max\left\{ t\in \mathcal{A}\::\: t\leq \tau_m \wedge \left((1-\tau_m) \frac{\alpha k}{m-k+1}\right) \right\},\:	k<m;
  			\end{aligned}$}  \\  \relax
  [A-DBH-SD]& $\tau^{\tiny \mbox{min}}_k = \max\{t\in \mathcal{A}\::\: t\leq \alpha k/ (m -(1-\alpha)k+1) \}$\\ \relax
  [DBR-$\lambda$] & $\tau^{\tiny \mbox{min}}_k = \max\left\{t\in \mathcal{A}\::\: t\leq \lambda \wedge \left((1-\lambda) \frac{\alpha k}{m-k+1}\right)  \right\}$ \\
 	\bottomrule \relax \\
\end{tabular}\\

Our current implementations of [DBH-SU] and [A-DBH-SU] first determine $\tau_m$ by searching for $\tau_m$ in $\mathcal{A} \cap [\tau^{\tiny \mbox{min}}_m,1]$ and then determine all other $\tau_k$ simultaneously using $\tau_k\ge \tau^{\tiny \mbox{min}}_1$ instead of $\tau_k\ge \tau^{\tiny \mbox{min}}_k$. We take this approach  for simplicity and performance reasons. The stepdown procedures only use the latter constraint. These lower bounds help to reduce the computational burden considerably.



\subsection{Transformed $p$-value} \label{subsec:TransPval}

If we are only interested in the set of rejected null hypotheses and not in the critical values, we can significantly speed-up the program by skipping the explicit computation of the critical values and by directly considering the transformed $p$-values:
\begin{align}\label{equ:transfpvalues}
p'_{k} = \xi_k (p_{(k)}),\:1\leq k \leq m,
\end{align}
where the functions $\xi_k(\cdot)$, defined in Table~\ref{tab:ListProceduresTransformations}, are such that $\tau_k$ is the inverse of $\xi_k$ at point $\alpha k/m$. 
Note that while the elements of $\{p_{(k)},1\leq k \leq m\}$ are ordered, this is not necessarily the case for the elements of $\{p'_{k},1\leq k \leq m\}$.
The following proposition is obvious.



\begin{prop}
	For each of the critical values listed in Table~\ref{tab:ListProceduresTransformations} we have for all $1\leq k \leq m$,
	\begin{align}
	  p_{(k)} \le \tau _k & \Longleftrightarrow  p'_{k} \le \alpha k /m, \label{eq:equiv:transformation}
	\end{align}
	where $p'_{k}$ are the transformed $p$-values defined by \eqref{equ:transfpvalues}.
\end{prop}

A consequence of \eqref{eq:equiv:transformation} is that the step-up and step-down cutoffs can be  computed by using only the transformed $p$-values and the BH critical values as follows:
 \begin{align*}
  \hat{k}_{SU}&=\max\{k\::\: p'_{k}\leq \alpha k/m \}\\
 \hat{k}_{SD}&=\max\{k\::\: \mbox{ for all } k'\leq k,\:p'_{k'}\leq \alpha k'/m \}.
 \end{align*}

Thus, all of the above methods can be interpreted as variant of the classical (SU or SD) BH procedure, for which each $p$-value has been suitably transformed to account for discreteness.


\begin{sidewaystable}[!p]
	\begin{tabular}{lll}
		\toprule
		Procedure & Critical values & Transformation\\ \hline
		\begin{tabular}{l}	[DBH-SU]\\ \\ \\ \\ \\$k<m$ \end{tabular}
		& {$\begin{aligned}[t]
			\tau_m &=\max\left\{t \in \mathcal{A}\::\: \frac{1}{m}\sum_{i=1}^m \frac{F_i\left(t\right)}{1-F_i\left(t\right)} \leq \alpha \right\}\nonumber\\
			\tau_k &= \max\left\{t \in \mathcal{A}\::\: t\leq \tau_m,\:\frac{1}{m}\sum_{i=1}^m \frac{F_i\left(t\right)}{1-F_i\left(\tau_m\right)} \leq \alpha k/m \right\}
			\end{aligned}$}
		& 
		{$\begin{aligned}[t]
			\xi_m(t) &= \frac{1}{m}\sum_{i=1}^m \frac{F_i\left(t\right)}{1-F_i\left(t\right)}\\
			\xi_k(t) &= 
					\begin{cases}
			      \frac{1}{m}\sum_{i=1}^m \frac{F_i\left(t\right)}{1-F_i\left(\tau_m\right)} & ,t \le \tau_m\\
			      1 & ,\text{else}\\
			    \end{cases}
\\
			\end{aligned}$}\\
		\midrule \relax
		[DBH-SD]
		& {$\begin{aligned}[t]
			\tau_k = \max\left\{ t\in \mathcal{A}\::\:  \frac{1}{m}\sum_{i=1}^m \frac{F_i\left(t\right)}{1-F_i\left(t\right)} \leq \alpha k/m \right\}
			\end{aligned}$}
		& {$\begin{aligned}[t]
			\xi_k (t) &= 
			\frac{1}{m}\sum_{i=1}^m \frac{F_i\left(t\right)}{1-F_i\left(t\right)}\\
			\end{aligned}$}\\
		\midrule \relax
	\begin{tabular}{l}	[A-DBH-SU]\\ \\ \\ \\ \\$k < m$ \end{tabular}
		& {$\begin{aligned}[t]
			\tau_m &=\max\left\{ t\in \mathcal{A}\::\: \frac{1}{m}\sum_{i=1}^m \frac{F_i\left(t\right)}{1-F_i\left(t\right)} \leq \alpha \right\}\nonumber\\
			\tau_k &= \max\left\{ t\in \mathcal{A}\::\: t \leq \tau_m,\:
			\sum_{\ell=1}^{m-k+1}
			\left(  \frac{F\left(t\right)}{1-F\left(\tau_m\right)}\right)_{(\ell)}
			\leq \alpha k\right\}
			\end{aligned}$}
		& {$\begin{aligned}[t]
  		\xi_m (t) &=  \frac{1}{m}\sum_{i=1}^m \frac{F_i\left(t\right)}{1-F_i\left(t\right)}\\
  		\xi_k (t) &= 		\begin{cases}
		    \frac{1}{m}  \sum_{\ell=1}^{m-k+1}
			 	\left(  \frac{F\left(t\right)}{1-F\left(\tau_m\right)}\right)_{(\ell)} & ,t \le \tau_m\\
			  1 & ,\text{else}\\
			\end{cases}
			\\
			\end{aligned}$}\\
		\midrule \relax
		[A-DBH-SD]
		& {$\begin{aligned}[t]
			\tau_k = \max\left\{ t\in \mathcal{A}\::\:   \sum_{\ell=1}^{m-k+1} \left(  \frac{F\left(t\right)}{1-F\left(t\right)}\right)_{(\ell)}\leq \alpha k\right\}
			\end{aligned}$}
		& {$\begin{aligned}[t]
			\xi_k (t) &= \frac{1}{m} \sum_{\ell=1}^{m-k+1} \left(  \frac{F\left(t\right)}{1-F\left(t\right)}\right)_{(\ell)}\\
			\end{aligned}$}\\
		\midrule \relax
			\begin{tabular}{l}	[DBR-$\lambda$]\\ \\ \\ \\ \\$k<m$ \end{tabular}
		& {$\begin{aligned}[t]
			\tau_m &=\max\left\{ t\in \mathcal{A}\::\: \left( F\left(t\right)\right)_{(1)} \leq  \left((1-\lambda) m\alpha\right) \wedge \lambda \right\}\: \\
			\tau_k &= \max\left\{ t\in \mathcal{A}\::\: \left( F\left(t\right)\right)_{(1)} \leq \lambda,\:  \sum_{\ell=1}^{m-k+1}
			\left(  F\left(t\right)\right)_{(\ell)}\leq \alpha k(1-\lambda)\right\}
			\end{aligned}$}
		& {$\begin{aligned}[t]
		\xi_m (t) &=
			\begin{cases}
	      \frac{\left( F\left(t\right)\right)_{(1)}}{m(1- \lambda)}  &\qquad ,\left( F\left(t\right)\right)_{(1)} \le \lambda\\
			  1 & \qquad ,\text{else}\\
			\end{cases}
			\\
			\xi_k (t) &=
			\begin{cases}
        \frac{\sum_{\ell=1}^{m-k+1}
				\left(  F\left(t\right)\right)_{(\ell)}}{m(1- \lambda)}   &\qquad ,\left( F\left(t\right)\right)_{(1)} \le \lambda\\
			  1 & \qquad ,\text{else}\\
			\end{cases}
			\\
			\end{aligned}$}\\
		\bottomrule\\
	\end{tabular}
	\caption{Implemented procedures (left column), critical values (center) and associated transformation functions (right column). The suffix 'SU' stands for step-up, the suffix 'SD' for step-down procedures.}
	\label{tab:ListProceduresTransformations}
\end{sidewaystable}



\subsection{Adjusted $p$-values} 

In applications, it is often convenient for the analyst to use \emph{adjusted} $p$-values $\widetilde{p}_k$ instead of the raw $p$-values and rejecting those hypotheses for which $\widetilde{p}_k \le \alpha$. The advantage of this approach is that it is more convenient to apply and easier to communicate. Furthermore, it avoids to explicitly rely on the, often somewhat arbitrary, choice of $\alpha$. With the  transformations introduced above, it is straightforward to define (variants of) discrete FDR-adjusted $p$-values. The generic definition given in \cite{DudoitLaan2007} then yields
\begin{align}
 \widetilde{p}_{k} &= \min_{\ell=k, \ldots,m } \left( \frac{m}{\ell} \cdot p'_{\ell} \right) \wedge 1,\:\:\: 1\leq k \leq m\label{adjustedpvaluesSU}
 \intertext{for step-up procedures and}
 \widetilde{p}_{k} &= \max_{\ell=1, \ldots,k } \left(\frac{m}{\ell} \cdot p'_{\ell} \right) \wedge 1,\:\:\: 1\leq k \leq m,\label{adjustedpvaluesSD}
\end{align}
for step-down procedures. 
For our step-down procedures, the usual result holds true.
\begin{prop}\label{prop:adjpvalues}
	For the step-down procedures [DBH-SD], [A-DBH-SD] and the step-up procedure [DBR-$\lambda$] listed in Table~\ref{tab:ListProceduresTransformations} we have for all $\alpha\in (0,1)$, for all $1\leq k \leq m$,
	\begin{align*}
	  \text{ $H_{(k)}$ is rejected by the procedure taken at level $\alpha$}  \:\:\:\Longleftrightarrow \:\:\: \widetilde{p}_{k} \le \alpha ,
	\end{align*}
	where $\widetilde{p}_{k}$ are the adjusted $p$-values defined by \eqref{adjustedpvaluesSD}.
\end{prop}
In the above proposition, note that $H_{(k)}$ is given by the original ordering of the $p$-values $\{p_i,1\leq i \leq m\}$.

For the procedures [DBH-SU], [A-DBH-SU],  the situation is more complicated since the adjusted $p$-value $\widetilde{p}_{k}$ depends on $\alpha$ (through $\tau_m$). The statement in Proposition~\ref{prop:adjpvalues} actually still holds in that situation but not the usual interpretation that the adjusted $p$-value $\widetilde{p}_{k}$ is the smallest level $\alpha$ at which the procedure rejects $H_{(k)}$. Hence, the analyst would need to exercise care in interpreting them. To avoid any confusion, the package does not report adjusted $p$-values for [DBH-SU] and [A-DBH-SU].

\section[Implementation]{Implementation in the package  \pkg{DiscreteFDR}} \label{sec:implementation}


\subsection{General structure} \label{subsec:General}

The package consists of four groups of functions:

\begin{longtable}{ll}
  Main functions    & \cDiscBH \\
                    & \cDBR \\
  Kernel functions  & \code{kernel.DBH.crit} \\
                    & \code{kernel.DBH.fast} \\
                    & \code{kernel.ADBH.crit} \\
                    & \code{kernel.ADBH.fast} \\
                    & \code{kernel.DBR.crit} \\
                    & \code{kernel.DBR.fast} \\
  Helper functions  & \code{match.pvals} \\
                    & \code{build.stepfuns} \\
                    & \code{short.eff} \\
                    & \code{fisher.pvalues.support} \\
  Wrapper functions & \cDBH \\
                    & \cADBH \\
                    & \code{fast.discrete} \\
\end{longtable}

The \cDiscBH function implements [DBH-SU], [DBH-SU], [A-DBH-SU] and [A-DBH-SD]. Similarly, \cDBR implements [DBR-$\lambda$]. They use the first three of the helper functions for common operations (see details in \ref{subsec:CompDetails}) and the kernels for the actual computation. [DBH-SU], [DBH-SD], [A-DBH-SU] and [A-DBH-SD] can be accessed directly through the wrapper functions \cDBH and \cADBH, respectively.


The wrapper function \code{fast.discrete} applies the Discrete FDR-controlling procedures, which are implemented in \cDiscBH, to a set of $2 \times 2$ contingency tables, given by a matrix or data frame. It uses the \code{fisher.pvalues.support} helper function to compute $p$-value c.d.f.s and raw $p$-values from these tables in the framework of Fisher's exact test.

We also provide the \code{amnesia} data set, used in our examples in Section \ref{sec:Using} and in our paper \citetalias{DDR2018}. It is basically the amnesia data set of package \pkg{discreteMTP}, but slightly reformatted (the difference lies in the third column).

The end user should only use the main functions \cDBR and \cDiscBH, and the wrapper functions \code{fast.discrete}, \cDBH and \cADBH. The other functions are only internal functions called by the main ones. We intentionally did not hide them, so that interested users would be able to understand how the main procedures work.

The functions \cDiscBH, \cDBH, \cADBH and \cDBR take the following input values:

\begin{longtable}{ll}
	\code{raw.pvalues}     & The vector (of the same length as \code{pCDFlist}) of raw observed $p$-values\\
	                       & which is calculated from the data.\\
	\code{pCDFlist}        & A list of vectors that represent the supports $\mathcal{A}_1, \ldots, \mathcal{A}_m$ of the discrete\\
	                       & distribution functions $F_1, \ldots, F_m$ under the respective null hypotheses,\\
	                       & as described in Section \ref{sec:mathematics}.\\
	\code{alpha}           & The global significance level $\alpha \in (0,1)$ at which the procedure provides\\
	                       & FDR control; the default is $0.05$.\\
	\code{direction}       & (\cDBH and \cADBH only) A string, either \code{"su"} or \code{"sd"}, specifying whether\\
	                       & the step-up variant (the default, \code{"su"}) or the step-down variant \\
	                       & (\code{"sd"}) should be used.\\
	\code{adaptive}        & Specifying whether the adaptive version is to be used (\code{TRUE}) or not\\
	                       & (\code{FALSE}).\\
	\code{lambda}          & (\cDBR only) The $\lambda$ parameter of the [DBR-$\lambda$] procedure as in Table \ref{tab:ListProceduresTransformations};\\
	                       & the default is $0.05$.\\
	\code{ret.crit.consts} & Specifying whether the critical values $\tau_k$ are to be computed and\\
                         & included in the output list, at the expense of computational speed; \\
                         & the default is \code{FALSE}.
\end{longtable}

They provide the following outputs:

\begin{longtable}{ll} 
	 \code{Rejected}           & A vector containing the rejected raw $p$-values\\
	 \code{Indices}            & A vector containing the indices of rejected hypotheses\\
	 \code{k.hat}              & Number of rejected hypotheses. This corresponds to $\hat{k}_{SU}$ and $\hat{k}_{SD}$,\\
	                           & as described in Section \ref{subsec:TransPval}.\\
	 \code{Alpha}              & Maximum significance level for the transformed $p$-values for which\\
	                           & a rejection occurred, that is \code{Alpha} $= \alpha \cdot$ \code{k.hat}$/ m$. This corre-\\
	                           & sponds to $\tau_k$ $(k = 1, \ldots, m)$ as in Section \ref{sec:mathematics}.\\
	 \code{Critical.constants} & A vector containing the critical values (if requested)\\
	 \code{Adjusted}           & A vector containing adjusted $p$-values (if available)\\
	 \code{Lambda}             & (\cDBR only) The parameter \code{lambda} that was used when calling \cDBR
\end{longtable}

More details as to the implementation are provided in the following part.

\subsection{Details for some specific functions} \label{subsec:CompDetails}

\subsubsection{Helper functions}

The \code{match.pvals} function performs nearest-neighbor matching for all elements of raw.pvalues, i.e., it checks for each value whether it occurs in its respective $p$-value c.d.f. If this is not the case, it is replaced by the value that is closest to it, that is, its nearest neighbor in its c.d.f. This is to ensure that all values of raw.pvalues actually originate from their respective c.d.f.s, e.g., to correct rounding errors. It has been inspired by a help page of the package \pkg{discreteMTP}.

\code{build.stepfuns} converts the vectors in \code{pCDFlist} to step function objects. This makes them easier to evaluate in the kernel functions. It is assumed and required that $F_i(t) \leq t$ applies to all c.d.f.s. Compliance with this premise cannot be not checked, so the user is responsible for providing correct vectors. If this condition is not met, the results may be incorrect.

The \code{short.eff} function extracts all values from a sorted vector that are greater than or equal to the effective critical value associated to a threshold. It simply replaces multiple recurring lines of code with one single function call.

\code{fisher.pvalues.support} computes discrete raw $p$-values and their support for the test of no association between two categorical variables in 2 x 2 contingency tables using Fisher's exact tests. The $p$-values are computed directly by \code{phyper}, instead of \code{fisher.test}, because the latter is much slower. The function is used by the \code{fast.discrete} function to apply such contingency tables directly to the \cDiscBH function.

\subsubsection{Main Functions}

Basically, both main functions have the same workflow:

\begin{enumerate}
  \item Use \code{match.pvals} for matching of raw $p$-values with the c.d.f.s and sort the results in ascending order.
  \item Convert the c.d.f. vectors in \code{pCDFlist} to step functions by means of \code{build.stepfuns}.
  \item Determine the overall support $\mathcal{A} = \bigcup \mathcal{A}_i$ from the individual c.d.f.s, remove double values and sort them in ascending order.
  \item Use the knowledge of lower bounds, as presented in Section \ref{subsec:CritConsts}, to remove unnecessary elements from the support.
  \item Compute transformed $p$-values and/or critical values (if requested) with kernel functions (see \ref{subsubsec:KernelFunctions}).
  \item Create output list with elements as described in Section \ref{subsec:General}.
\end{enumerate}

\subsubsection{Kernel functions} \label{subsubsec:KernelFunctions}

As stated in Section \ref{sec:mathematics}, there are two ways to determine which hypotheses corresponding to the elements of a raw $p$-value vector can be rejected.
\begin{enumerate}
  \item With critical values (see \ref{subsec:CritConsts}): this approach works by first determining the critical values. Especially when the size of the support and the number of hypotheses are large, it is computationally intensive, because all elements of the support have to be evaluated by every single c.d.f.
  \item With transformed $p$-values (see Section \ref{subsec:TransPval}): here, only the raw $p$-values are evaluated. Thus, it is much more efficient.
\end{enumerate}
As a result, all three implemented procedures have two kernels, that is, a fast one for simplified computation and a slower implementation that calculates critical values. These values are then used to determine which hypotheses are to be rejected and which are not.

The kernel functions need the following parameters:

\begin{longtable}{ll}
  \code{stepf}    & A list of step function objects.\\
  \code{pv.numer} & A vector of values from the support for the argument of the $F_i$ in the\\
                  & numerators of the fractions in the formulas presented in Table \ref{tab:ListProceduresTransformations}.\\
  \code{pv.denom} & A vector of $p$-values or a single one for the denominators as in Table \ref{tab:ListProceduresTransformations}.
\end{longtable}

The critical values kernels additionally need:

\begin{longtable}{ll}
  \code{alpha}     & A numeric value specifying the global significance level.\\
  \code{sorted.pv} & A numeric vector of observed $p$-values in ascending order.
\end{longtable}

For the [DBH] and [ADBH] procedures, a direction (\code{"su"} for step-up or \code{"sd"} for step-down) do not need to be explicitly passed to the kernels, because these two cases can be distinguished by \code{pv.numer} (the input values for the numerators of the fractions presented in Table \ref{tab:ListProceduresTransformations}) and \code{pv.denom} (the value(s) for the respective denominators). If all elements of both vectors are identical, we have the step-down case. Otherwise, it is step-up.

Basically, the kernels implement the formulas of Table \ref{tab:ListProceduresTransformations}. Every step function must be evaluated at every element of either the support (for the critical values approach) or only the sorted raw $p$-values (for the transformed $p$-values approach). If this were done by means of \code{sapply} and/or \code{apply}, the results would be stored in a matrix, for which enough memory is reserved automatically. For a large number of hypotheses and even larger support sets, the size of this matrix would be vast and may easily take up several, if not dozens, of gigabytes of RAM. This may be too much for many workstations. As a solution to this problem, we implemented memory-conserving algorithms.  

For [DBH-SD], this means, that, for each $p$-value c.d.f. $F_i$, we compute the fractions $\frac{F_i(t)}{1 - F_i(t)}$ for all of $\{t \in \mathcal{A} : t \geq \tau^{\tiny \mbox{min}}_1\}$, with $\alpha$ being the significance level (see \citetalias{DDR2018}, Lemma 3) inside a \code{for} loop, which adds up the resulting \proglang{R} vectors iteratively. If the critical values are not demanded by the user, we evaluate at the observed $p$-values instead of the support. In both cases, the number of passes of the \code{for} loop is identical to the number of hypotheses. For [DBH-SU], we first compute the (last) critical constant $\tau_m$ as above, but we can restrict the computations to the set $\{t \in \mathcal{A} : t \geq \frac{\alpha}{1 + \alpha}\}$ (see \citetalias{DDR2018}, Lemma 2). After that, we compute the fractions $\frac{F_i(t)}{1 - F_i(\tau_m)}$ as before, but we only have to consider values of the set $\{t \in \mathcal{A} : t \leq \tau_m\}$.

For the [A-DBH] procedures, the step functions are evaluated iteratively at smaller chunks of the input vectors \code{pv.numer} and \code{pv.denom}. The results of the fractions are then stored in a matrix. We found a size of 256 MiB to deliver the best performance. Depending on the number of hypotheses, $m$, the size and number of the chunks is determined dynamically. All $p$-value transformations and critical constant computations are done for this submatrix. The intermediate results are then stored in vectors. This is repeated for the remaining chunks by using a \code{for} loop. The intermediate results are updated with each pass of the loop until all input values have been processed. The [DBR-$\lambda$] algorithm is working almost identically, but no fractions are needed and there is no step-up/step-down direction.

\subsection{Run times}

To illustrate the run times of \cDBH, \cADBH and \cDBR, we used the \code{arabidopsisE} data set, which was once included in the \pkg{fdrDiscreteNull} package, but was removed in recent versions. From this data, a total of 17400 hypotheses along with their respective $p$-value distributions and a vector of raw $p$-values were derived (for more details, see \cite{fdrDiscreteNull}). The accumulated size of the support $\mathcal{A}$ is 1,074,398. From this data, we used subsets of the first $m = 250, 500, 1000, 3000, 5000, 7000, 10000, 17400$ hypotheses, each resulting in different support sizes, as shown in the tables in the appendix. For each subset, the median run time of 25 runs was recorded. The decision for multiple, repeated runs and their median was made in order to account for possible side loads of the workstation and to avoid overly pronounced effects of very good and especially very bad runs, so we get a robust indication of the required time. All three methods were used with the following settings:
\begin{itemize}
  \item \code{alpha = 0.05}
  \item \code{direction = "sd"} and \code{direction = "su"} (\cDBH and \cADBH only)
  \item \code{lambda = 0.05} (\cDBR only)
  \item \code{ret.crit.consts = TRUE} and \code{ret.crit.consts = FALSE}
\end{itemize}
All computations were performed with \proglang{R} version 3.5.1 on the following system:
\begin{itemize}
  \item CPU: AMD Ryzen 7 1800X, 3.60 GHz
  \item RAM: 32 GiB DDR4, 2400 MHz
  \item OS: Windows 10 Education v1803
\end{itemize}

The complete results tables can be found in the appendix.

\subsubsection{Results of critical values approach}

The following plots illustrate our findings by depicting the development of the run times as a function of the product of $m$ and the overall support size $|\mathcal{A}|$. In addition to a plot with standard axis scaling, we also employ an additional one with logarithmic axes.

\begin{figure}[H]
\centering
\includegraphics{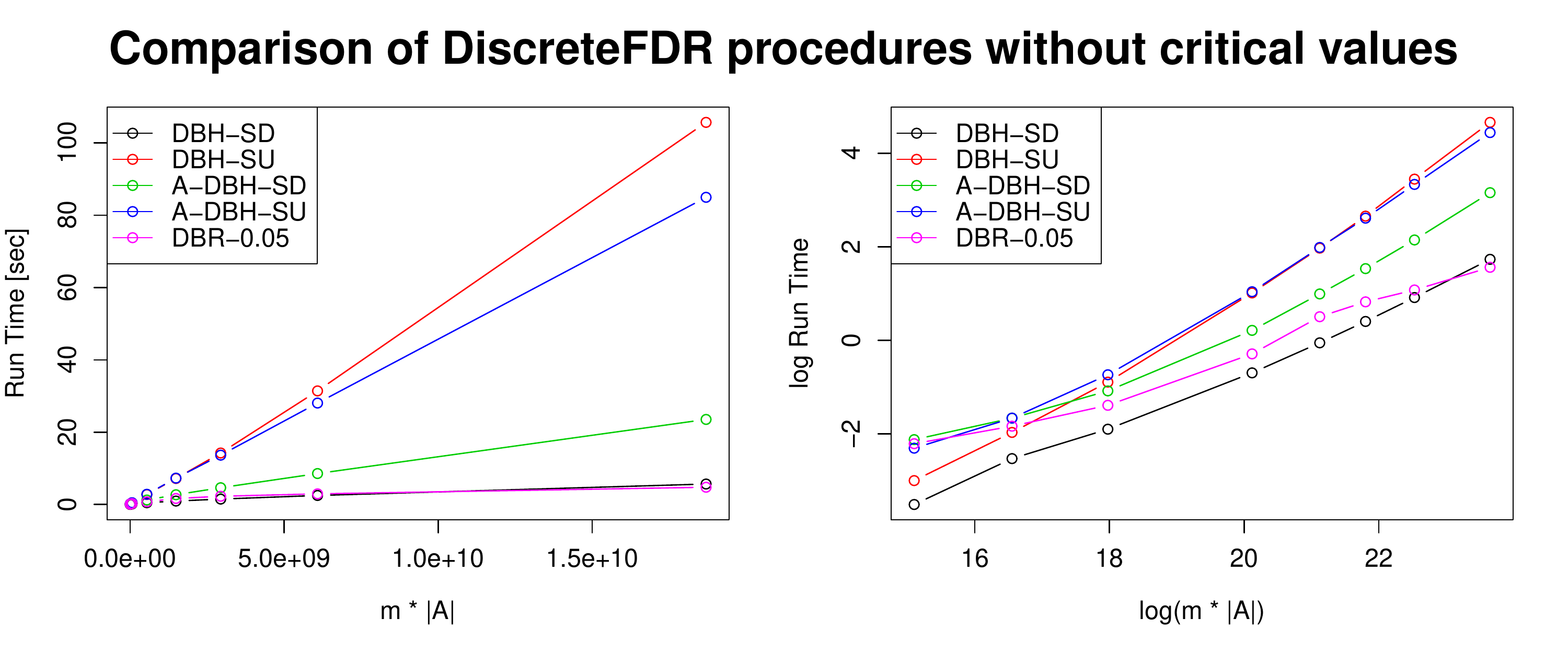}
\caption{Run time comparison of \pkg{DiscreteFDR} procedures with computation of critical values.}
\end{figure}

From both plots, we can clearly observe that [DBH-SU] is the fastest algorithm, followed by [DBH-SD], whose computation takes about 1.5 times as long. The calculations of [A-DBH-SU] takes about 4 times and those of [A-DBH-SD] almost 7 times as long as [DBH-SU]. [DBR-0.05] needs almost exactly the same time as [A-DBH-SU], so that their respective lines in the plots overlap. In addition, the second plot shows that these proportions of run times and, as a result, the order remain stable after $m \cdot |\mathcal{A}| \approx 5,000 \cdot 300,000 = 1,500,000,000$. Furthermore, it is visible that the run times of all the procedures exhibit roughly linear growth.

\subsubsection{Results of transformed $p$-values approach}

\begin{figure}[H]
\centering
\includegraphics{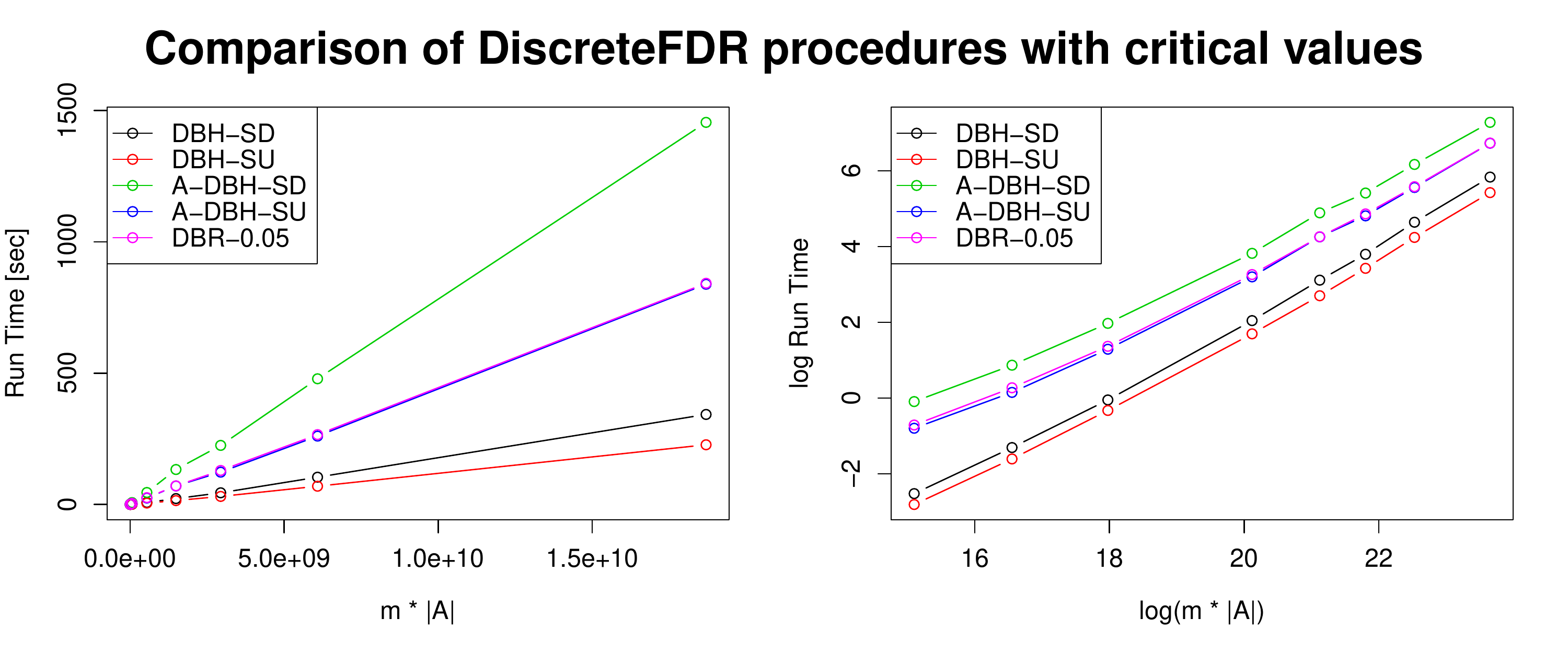}
\caption{Run time comparison of \pkg{DiscreteFDR} procedures without computation of critical values.}
\end{figure}

Here, it is immediately apparent that the transformed $p$-values approach is an order of magnitude faster than the ones with critical values, but recognizing a ranking is a bit more difficult. However, up to and including $m \cdot |\mathcal{A}| \approx 1,000 \cdot 64,000 = 64,000,000$, all procedures take less than a second to compute their results, which is almost unnoticable. After that point, [DBH-SD] and [DBR-0.05] are the fastest algorithms, with [DBR-0.05] outperforming every other procedure for very large sizes. They are followed by [A-DBH-SD], [A-DBH-SU] and [DBH-SU]. The two latter ones exhibit mostly identical performance, but the higher $m \cdot \mathcal{A}$, the higher the performance advantage of [A-DBH-SU] over [DBH-SU], although they remain the slowest methods. Their largely higher computation time is explained by the fact that these two procedures have to determine the critical value $\tau_m$, which is responsible for ~80\% of the computational time, as an in-depth analysis has shown. The increasing advantage of [A-DBH-SU] over [DBH-SU] is explained by the fact that, as described before, [DBH-SU] simply adds up the fractions of evaluated c.d.f.s with a for loop, while [A-DBH-SU] uses a chunking approach, which also uses \code{for} loops, but requires much fewer passes than [DBH-SU]. The advantage of this approach is mitigated by the required sorting. But sill, as a result, [DBH-SU] becomes less efficient with larger sizes of $m \cdot \mathcal{A}$.

%
%

\section[Examples]{Further analyses} \label{sec:Using}
\subsection{Analysis of pharmacovigilance data}
To illustrate how the procedures in \pkg{DiscreteFDR} can be used for real data, we revisit the analysis of the pharmacovigilance data from \citet{Heller2012} performed in \citetalias{DDR2018}. This data set is obtained from a database for reporting, investigating and monitoring
adverse drug reactions due to the Medicines and Healthcare products Regulatory Agency in the United Kingdom. It contains the number of reported cases of amnesia as well as the total number of adverse events reported for each of the $m = 2446$ drugs in the database. For more details we refer to \citet{Heller2012} and to the accompanying R-package \pkg{discreteMTP} (\cite{discreteMTP}), which also contains the data. \cite{Heller2012} investigate the association between reports of amnesia and suspected drugs by performing for each drug a Fisher's exact test (one-sided) for testing association between the drug
and amnesia while adjusting for multiplicity by using several (discrete) FDR procedures. In what follows we present code that reproduces parts of Figure 2 and Table 1 in  \citetalias{DDR2018}.

We procede as in the example in section \ref{sec:ToyExample}. Since we need to access the critical values  we first determine the $p$-values and their support for the data set \code{amnesia} contained for convenience in the package \pkg{DiscreteFDR}. For this, we use the option \code{"HG2011"} in the function \code{fisher.pvalues.support}.

\begin{Schunk}
\begin{Sinput}
R> library("DiscreteFDR")
R> data(amnesia)
R> amnesia.formatted <- fisher.pvalues.support(amnesia[, 2:3], 
+                                              input = "HG2011")
R> raw.pvalues <- amnesia.formatted$raw
R> pCDFlist <- amnesia.formatted$support
\end{Sinput}
\end{Schunk}

Then we perform the FDR analysis with functions \code{DBH} and \code{ADBH} (SU and SD) and \code{DBR} at level $\alpha=0.05$ including critical values.
\begin{Schunk}
\begin{Sinput}
R> DBH.su <- DBH(raw.pvalues, pCDFlist, ret.crit.consts = TRUE)
R> DBH.sd <- DBH(raw.pvalues, pCDFlist, direction = "sd", 
+                ret.crit.consts = TRUE)
R> ADBH.su <- ADBH(raw.pvalues, pCDFlist, ret.crit.consts = TRUE)
R> ADBH.sd <- ADBH(raw.pvalues, pCDFlist, direction = "sd", 
+                  ret.crit.consts = TRUE)
R> DBR.su <- DBR(raw.pvalues, pCDFlist, ret.crit.consts = TRUE)
\end{Sinput}
\end{Schunk}

By accessing the critical values we can now generate a plot similar to Figure 2 from \citetalias{DDR2018}. Note that both [DBH-SU] and [DBH-SD] are visually indistinguishable from their adaptive counterparts.

%

\begin{figure}[H]
\centering
\includegraphics{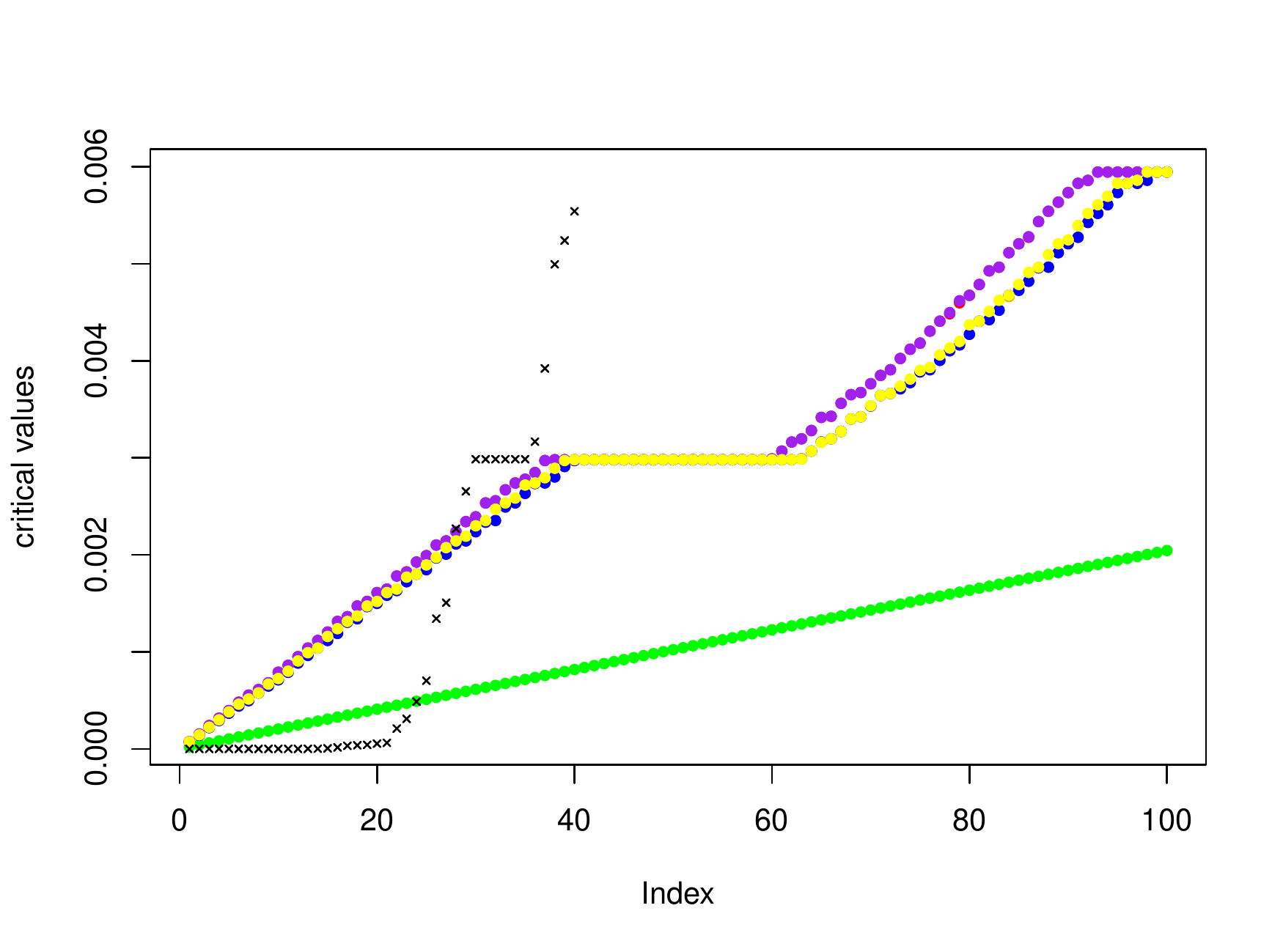}
\caption{\label{fig:anna} Critical values for [BH] (green dots), [DBH-SU] (orange dots), [DBH-SD] (red dots), [A-DBH-SU] (blue dots), [A-DBH-SD] (purple dots), [DBR] (yellow dots). The sorted raw $p$-values are represented by asterisks.}
\end{figure}

The rejected hypotheses can be accessed via the command \code{$Indices}. The following code yields some of the values  from Table 1 in \citetalias{DDR2018}:

\begin{Schunk}
\begin{Sinput}
R> rej.BH <- length(which(p.adjust(raw.pvalues, method = "BH") <= 0.05))
R> rej.DBH.su <- length(DBH.su$Indices)
R> rej.DBH.sd <- length(DBH.sd$Indices)
R> rej.ADBH.su <- length(ADBH.su$Indices)
R> rej.ADBH.sd <- length(ADBH.sd$Indices)
R> rej.DBR.su <- length(DBR.su$Indices)
R> c(rej.BH, rej.DBH.su, rej.DBH.sd, rej.ADBH.su, rej.ADBH.sd, rej.DBR.su)
\end{Sinput}
\begin{Soutput}
[1] 24 27 27 27 27 27
\end{Soutput}
\end{Schunk}
The (continuous) BH rejects only 24 hypotheses whereas all the discrete procedures implemented in \pkg{DiscreteFDR} are able to identify three additional drug candidates potentially associated with amnesia.

\subsection{Other types of discrete tests} \label{ssec:PoissonTests}
In this section we sketch how \pkg{DiscreteFDR} can be used to analyse arbitrary multiple discrete tests. \cite{JimenezUnaAlvarez2018} used  \pkg{DiscreteFDR} to detect disorder in NGS experiments based on one-sample tests of the Poisson mean. Rather than reproducing their analysis in detail, we illustrate the general approach by using a toy example similar to the one in Section \ref{sec:ToyExample} and show how the test of the Poisson mean can be accomodated by \pkg{DiscreteFDR}. 

To fix ideas, suppose we observe $m=9$ independent Poisson distributed counts $N_1, \ldots, N_9$  (\cite{JimenezUnaAlvarez2018} used this to model the read counts of different DNA bases). We assume that $N_i \sim \PoisVert (\lambda_i)$ and the goal is to identify cases where $\lambda_i$ is larger than some pre-specified value $\lambda^0_i$, i.e., we have the (one-sided) multiple testing problem
\begin{align*}
H_{0i}: \lambda_i = \lambda^0_i & \qquad \text{vs.} \qquad H_{1i}: \lambda_i > \lambda^0_i.
\end{align*}
As in Section \ref{sec:ToyExample}, the goal is to adjust for multiple testing by using the [DBH-SD] procedure at FDR-level $\alpha=5\%$. In our example the observations $n_1,\ldots, n_9$ and parameters $\lambda^0_1, \ldots, \lambda^0_9$ are given as follows:
\begin{Schunk}
\begin{Soutput}
      observations lambda.0
 [1,]            3      0.6
 [2,]            3      1.2
 [3,]            1      0.7
 [4,]            2      1.3
 [5,]            3      1.0
 [6,]            3      0.2
 [7,]            1      0.8
 [8,]            2      1.3
 [9,]            4      0.9
\end{Soutput}
\end{Schunk}
Denote by $G_i$ the distribution of $N_i$ under $H_{0i}$  i.e., $G_i(x)=P(N_i \le x)$. For observations $n_1,\ldots, n_9$ of $N_1, \ldots,N_9$ the $p$-values for the above one-sided test are given by
\begin{align*}
p_i &= P(N_i \ge n_i) =P(N_i > n_i-1)= \overline{G_i}(n_i-1), 
\end{align*}
where $\overline{G_i}(x)=P(N_i >x)=1-G_i(x)$ denotes the survival function of the Poisson distribution with parameter $\lambda^0_i$. Thus the  raw $p$-values are determined by the following \proglang{R} code:
\begin{Schunk}
\begin{Sinput}
R> raw.pvalues <- sapply(1:m,function(i){ppois(observations[i]-1,lambda.vector[i], 
+                                              lower.tail = FALSE)})
R> raw.pvalues
\end{Sinput}
\begin{Soutput}
[1] 0.023115288 0.120512901 0.503414696 0.373176876 0.080301397
[6] 0.001148481 0.550671036 0.373176876 0.013458721
\end{Soutput}
\end{Schunk}
Following the definition of the \code{qpois} function in \proglang{R} we define the inverse function of $\overline{G_i}$ by
\begin{align*}
\overline{G_i}^{-1}(p) &= \min \{ n \in \nat : \overline{G_i}(n) \le  p \} \\
\intertext{and obtain for the distribution function of the $i$-th $p$-value under the null}
F_i(x) &= \overline{G_i} ( \overline{G_i}^{-1}(x) ).
\end{align*}
Each function $F_i$ is a step function with $F_i(0)=0$, $F_i(1)=1$ and there exists an infinite sequence of jumps at locations $1=x_1 > x_2 > \ldots > x_n > x_{n+1}> \ldots > 0$ such that $F(x_j)=x_j$ for $j \in \nat$.

Initially it seems that we run into a problem if we want to determine the critical values of [DBH-SD] since the supports of $F_1, \ldots, F_9$ are no longer finite (but still discrete). We can deal with this problem by using the observation from Section \ref{subsec:CritConsts} that it is sufficient to consider new, restricted supports $\mathcal{A}_i \cap [s^{\tiny \mbox{min}},1]$ where the lower threshold satisfies
\begin{align}
s^{\tiny \mbox{min}} &\le \tau^{\tiny \mbox{min}}_1  =\max \left\{ t\in \mathcal{A}\::\: t\leq y^{\tiny \mbox{min}}  \right\} \qquad \text{where} \qquad y^{\tiny \mbox{min}} = \frac{\alpha}{m} \cdot \left( 1+\frac{\alpha}{m} \right)^{-1}. \label{eq:tau.min.Poisson}
\end{align}
To determine such an $s^{\tiny \mbox{min}}$ we procede as follows. Define $n^{\tiny \mbox{max}}_i = \overline{G_i}^{-1}(y^{\tiny \mbox{min}})+1,$ $
t^{\tiny \mbox{min}}_i = \overline{G_i} (n^{\tiny \mbox{max}}_i-1)$ and set $s^{\tiny \mbox{min}} = \min \left( t^{\tiny \mbox{min}}_1, \ldots, t^{\tiny \mbox{min}}_9 \right)$. It is easily checked that this choice of $s^{\tiny \mbox{min}}$ satisfies \eqref{eq:tau.min.Poisson}. We can determine  $s^{\tiny \mbox{min}}$ by the following code
\begin{Schunk}
\begin{Sinput}
R> y.min <- alpha/m*(1+alpha/m)^(-1)
R> n.max <- sapply(1:m,function(w){qpois(y.min,lambda.vector[w],
+        lower.tail = FALSE)})+1
R> t.min <- sapply(1:m,function(w){ppois(n.max[w]-1,lambda.vector[w],
+                                        lower.tail = FALSE)})
R> s.min <- min(t.min)
R> s.min
\end{Sinput}
\begin{Soutput}
[1] 0.0007855354
\end{Soutput}
\end{Schunk}

\begin{figure}[htb]
\centering
\includegraphics{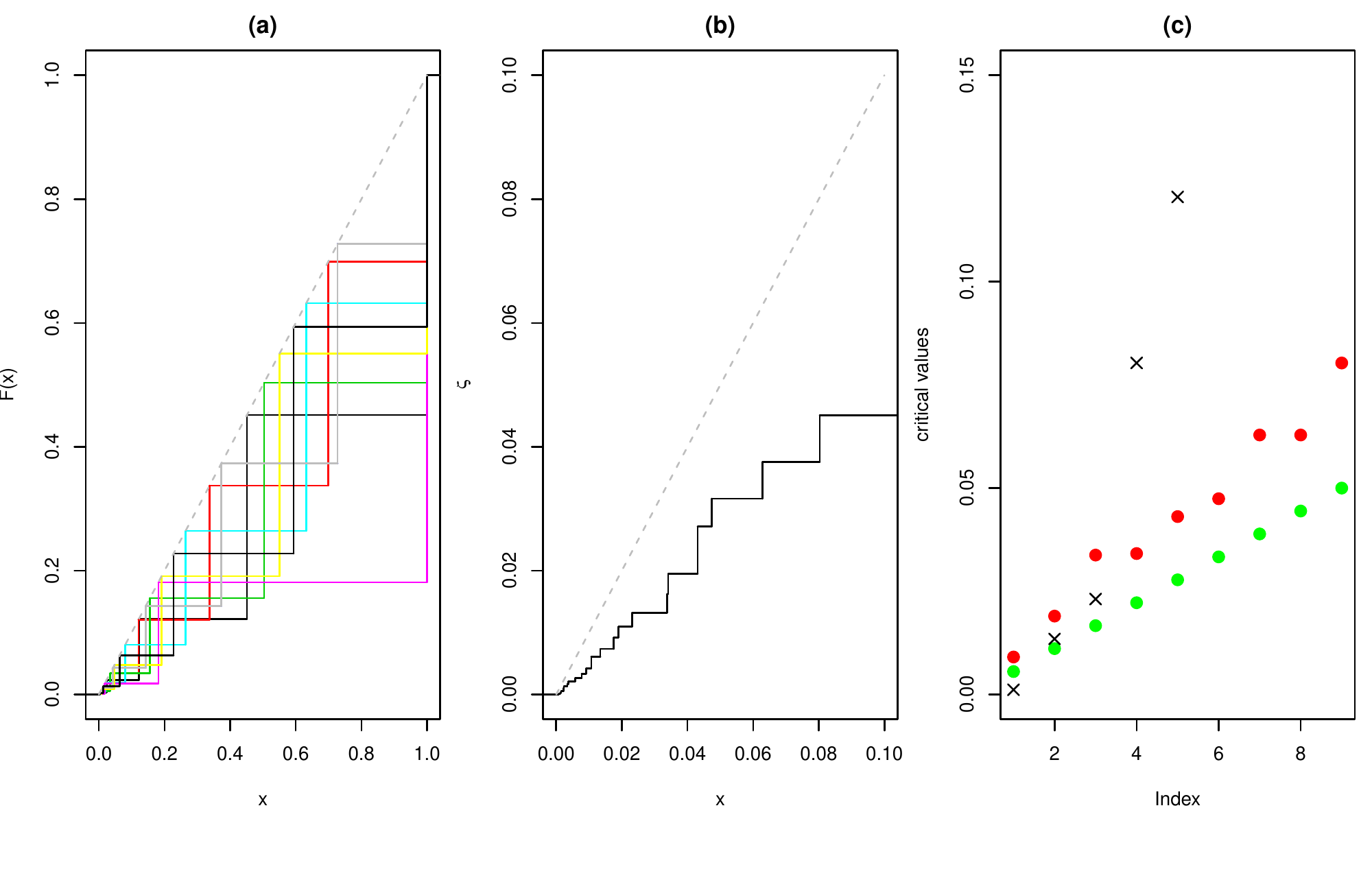}
\caption{\label{fig:Poisson} Panel (a) depicts the distribution functions $F_1, \ldots, F_9$ in various colours, (b) is a graph of the transformation function $\xiSD$. The uniform distribution function is shown in light grey in (a) and (b). Panel (c) shows the [BH] critical values (green dots), the DBH step-down critical values (red dots) and the sorted raw $p$-values (asterisks).}
\end{figure}

For determining the restricted supports it is actually more convenient to work with $n^{\tiny \mbox{max}}_i$ than $s^{\tiny \mbox{min}}$. We can subsequently use these supports as the \code{pCDFlist} argument in the usual way when calling the \code{DBH} function:

\begin{Schunk}
\begin{Sinput}
R> supports <- lapply(1:m,function(w){sort(ppois(0:n.max[w]-1,lambda.vector[w], 
+                                                lower.tail = FALSE))})
R> DBH.sd <- DBH(raw.pvalues,supports,direction = "sd", ret.crit.consts = TRUE)
\end{Sinput}
\end{Schunk}
Figure \ref{fig:Poisson} shows a  summary similar to Figure \ref{fig:otto}. Applying the continuous BH procedure 
\begin{Schunk}
\begin{Sinput}
R> p.adjust(raw.pvalues, method = "BH")
\end{Sinput}
\begin{Soutput}
[1] 0.06934586 0.21692322 0.55067104 0.47979884 0.18067814 0.01033633
[7] 0.55067104 0.47979884 0.06056424
\end{Soutput}
\end{Schunk}
results in one rejection at FDR-level $\alpha=5\%$, whereas the DBH step-up procedure can reject three hypotheses:
\begin{Schunk}
\begin{Sinput}
R> DBH.sd$Adjusted
\end{Sinput}
\begin{Soutput}
[1] 0.039602625 0.101622881 0.580898946 0.522450788 0.101509307
[6] 0.001935955 0.626257875 0.522450788 0.033073393
\end{Soutput}
\end{Schunk}
As in Figure \ref{fig:otto},  Panel (c) presents a graphical comparison between the two procedures applied to the $p$-values.

\section[Summary]{Summary and future work} \label{sec:summary}
Controlling the FDR for discrete tests is an important goal in many data analytic settings. In this paper, we introduced the \proglang{R} package \pkg{DiscreteFDR}, implementing procedures from \citetalias{DDR2018}. These procedures come with guaranteed FDR control under independence and deal effectively with the conservativeness encountered in discrete tests. 

We hope that our software will make discrete methods for FDR control more accessible to a wide audience of practitioners. More specifically, \pkg{DiscreteFDR} can be used both in an 'expert' and a 'standard' mode. For the data analyst, taking discreteness and multiplicity issues into account simultaneously may appear to be rather challenging since information on many distribution functions has to be stored, combined and evaluated. For this reason, we have included the wrapper function \code{fast.discrete} which applies the discrete procedures to a set of 2 $\times$ 2 tables, given by a matrix or data frame, where each contingency table is analysed by Fisher's exact test. Thus, this function can be seen as an implementation of a multiple Fisher test that controls FDR. For controlling the more stringent Familywise Error Rate (FWER) for multiple exact Fisher tests, we would like to point out the \proglang{R} package \pkg{multfisher} which implements the approaches described in \cite{Ristl2018}. For those analysts who are looking for a simple to apply, off-the-shelf method, using \code{fast.discrete} will automatically take care of generating the list of (the support of the) distribution functions \code{pCDFlist}, which may otherwise be tedious work. For more expert users who want to use other tests than Fisher's exact test, the work flow, described in more detail in Section \ref{ssec:PoissonTests}, consists of first generating the \code{pCDFlist} list, and then passing this on to the \code{DBH} or \code{DBR} functions.

Interfaces that generate \code{pCDFlist} from a given data set for a given statistical test are very helpful tools. Currently, Fisher's exact test is the only test for which our package supplies such an interface. In the future, we are planning to include helper functions similar to \code{fisher.pvalues.support} for further tests like the  Binomial and Poisson tests.


The \proglang{R} package \pkg{DiscreteFDR} is available from the Comprehensive \proglang{R} Archive network (CRAN) at \url{https://cran.r-project.org/web/packages/DiscreteFDR/index.html}.

\section[Acknowledgements]{Acknowledgements}
We thank Antje Jahn for very carefully reading the manuscript and providing  numerous suggestions that greatly improved the content and presentation of the paper. This work has been supported by the CNRS (PEPS FaSciDo) and the French grants ANR-16-CE40-0019 (SansSouci project) and ANR-17-CE40-0001 (Basics project).
\bibliography{biblio}

\begin{appendix}

\section{Run Time Comparison Tables} \label{apx:runTimeTables}

\begin{table}[H]
  \centering
  \begin{tabular}{r|r|r|r}
    $m$ & $|\mathcal{A}|$ & Critical values & Run time\\
    \hline
    \hline
    \multirow{2}{*}{250} & \multirow{2}{*}{14442} & \code{TRUE} & 0.08\\
    & & \code{FALSE} & 0.03\\
    \hline
    \multirow{2}{*}{500} & \multirow{2}{*}{30873} & \code{TRUE} & 0.27\\
    & & \code{FALSE} & 0.08\\
    \hline
    \multirow{2}{*}{1000} & \multirow{2}{*}{64058} & \code{TRUE} & 0.95\\
    & & \code{FALSE} & 0.15\\
    \hline
    \multirow{2}{*}{3000} & \multirow{2}{*}{181801} & \code{TRUE} & 7.73\\
    & & \code{FALSE} & 0.50\\
    \hline
    \multirow{2}{*}{5000} & \multirow{2}{*}{297930} & \code{TRUE} & 22.51\\
    & & \code{FALSE} & 0.95\\
    \hline
    \multirow{2}{*}{7000} & \multirow{2}{*}{420162} & \code{TRUE} & 44.59\\
    & & \code{FALSE} & 1.50\\
    \hline
    \multirow{2}{*}{10000} & \multirow{2}{*}{608459} & \code{TRUE} & 104.00\\
    & & \code{FALSE} & 2.50\\
    \hline
    \multirow{2}{*}{17400} & \multirow{2}{*}{1074398} & \code{TRUE} & 342.88\\
    & & \code{FALSE} & 5.66\\
    \hline
  \end{tabular}
  \caption{Median run times for the [DBH-SD] procedure.}
\end{table}


\begin{table}[H]
  \centering
  \begin{tabular}{r|r|r|r}
    $m$ & $|\mathcal{A}|$ & Critical values & Run time\\
    \hline
    \hline
    \multirow{2}{*}{250} & \multirow{2}{*}{14442} & \code{TRUE} & 0.06\\
    & & \code{FALSE} & 0.05\\
    \hline
    \multirow{2}{*}{500} & \multirow{2}{*}{30873} & \code{TRUE} & 0.20\\
    & & \code{FALSE} & 0.14\\
    \hline
    \multirow{2}{*}{1000} & \multirow{2}{*}{64058} & \code{TRUE} & 0.72\\
    & & \code{FALSE} & 0.41\\
    \hline
    \multirow{2}{*}{3000} & \multirow{2}{*}{181801} & \code{TRUE} & 5.45\\
    & & \code{FALSE} & 2.76\\
    \hline
    \multirow{2}{*}{5000} & \multirow{2}{*}{297930} & \code{TRUE} & 14.91\\
    & & \code{FALSE} & 7.19\\
    \hline
    \multirow{2}{*}{7000} & \multirow{2}{*}{420162} & \code{TRUE} & 30.76\\
    & & \code{FALSE} & 14.24\\
    \hline
    \multirow{2}{*}{10000} & \multirow{2}{*}{608459} & \code{TRUE} & 69.60\\
    & & \code{FALSE} & 31.41\\
    \hline
    \multirow{2}{*}{17400} & \multirow{2}{*}{1074398} & \code{TRUE} & 227.41\\
    & & \code{FALSE} & 105.66\\
    \hline
  \end{tabular}
  \caption{Median run times for the [DBH-SU] procedure.}
\end{table}


\begin{table}[H]
  \centering
  \begin{tabular}{r|r|r|r}
    $m$ & $|\mathcal{A}|$ & Critical values & Run time\\
    \hline
    \hline
    \multirow{2}{*}{250} & \multirow{2}{*}{14442} & \code{TRUE} & 0.91\\
    & & \code{FALSE} & 0.12\\
    \hline
    \multirow{2}{*}{500} & \multirow{2}{*}{30873} & \code{TRUE} & 2.38\\
    & & \code{FALSE} & 0.19\\
    \hline
    \multirow{2}{*}{1000} & \multirow{2}{*}{64058} & \code{TRUE} & 7.19\\
    & & \code{FALSE} & 0.34\\
    \hline
    \multirow{2}{*}{3000} & \multirow{2}{*}{181801} & \code{TRUE} & 45.73\\
    & & \code{FALSE} & 1.24\\
    \hline
    \multirow{2}{*}{5000} & \multirow{2}{*}{297930} & \code{TRUE} & 133.17\\
    & & \code{FALSE} & 2.70\\
    \hline
    \multirow{2}{*}{7000} & \multirow{2}{*}{420162} & \code{TRUE} & 225.01\\
    & & \code{FALSE} & 4.64\\
    \hline
    \multirow{2}{*}{10000} & \multirow{2}{*}{608459} & \code{TRUE} & 478.63\\
    & & \code{FALSE} & 8.56\\
    \hline
    \multirow{2}{*}{17400} & \multirow{2}{*}{1074398} & \code{TRUE} & 1454.77\\
    & & \code{FALSE} & 23.52\\
    \hline
  \end{tabular}
  \caption{Median run times for the [A-DBH-SD] procedure.}
\end{table}


\begin{table}[H]
  \centering
  \begin{tabular}{r|r|r|r}
    $m$ & $|\mathcal{A}|$ & Critical values & Run time\\
    \hline
    \hline
    \multirow{2}{*}{250} & \multirow{2}{*}{14442} & \code{TRUE} & 0.45\\
    & & \code{FALSE} & 0.10\\
    \hline
    \multirow{2}{*}{500} & \multirow{2}{*}{30873} & \code{TRUE} & 1.16\\
    & & \code{FALSE} & 0.19\\
    \hline
    \multirow{2}{*}{1000} & \multirow{2}{*}{64058} & \code{TRUE} & 3.63\\
    & & \code{FALSE} & 0.48\\
    \hline
    \multirow{2}{*}{3000} & \multirow{2}{*}{181801} & \code{TRUE} & 24.54\\
    & & \code{FALSE} & 2.83\\
    \hline
    \multirow{2}{*}{5000} & \multirow{2}{*}{297930} & \code{TRUE} & 70.62\\
    & & \code{FALSE} & 7.29\\
    \hline
    \multirow{2}{*}{7000} & \multirow{2}{*}{420162} & \code{TRUE} & 123.09\\
    & & \code{FALSE} & 13.64\\
    \hline
    \multirow{2}{*}{10000} & \multirow{2}{*}{608459} & \code{TRUE} & 260.41\\
    & & \code{FALSE} & 28.05\\
    \hline
    \multirow{2}{*}{17400} & \multirow{2}{*}{1074398} & \code{TRUE} & 838.53\\
    & & \code{FALSE} & 84.94\\
    \hline
  \end{tabular}
  \caption{Median run times for the [A-DBH-SU] procedure.}
\end{table}


\begin{table}[H]
  \centering
  \begin{tabular}{r|r|r|r}
    $m$ & $|\mathcal{A}|$ & Critical values & Run time\\
    \hline
    \hline
    \multirow{2}{*}{250} & \multirow{2}{*}{14442} & \code{TRUE} & 0.49\\
    & & \code{FALSE} & 0.11\\
    \hline
    \multirow{2}{*}{500} & \multirow{2}{*}{30873} & \code{TRUE} & 1.31\\
    & & \code{FALSE} & 0.16\\
    \hline
    \multirow{2}{*}{1000} & \multirow{2}{*}{64058} & \code{TRUE} & 3.92\\
    & & \code{FALSE} & 0.25\\
    \hline
    \multirow{2}{*}{3000} & \multirow{2}{*}{181801} & \code{TRUE} & 26.10\\
    & & \code{FALSE} & 0.75\\
    \hline
    \multirow{2}{*}{5000} & \multirow{2}{*}{297930} & \code{TRUE} & 70.89\\
    & & \code{FALSE} & 1.66\\
    \hline
    \multirow{2}{*}{7000} & \multirow{2}{*}{420162} & \code{TRUE} & 129.52\\
    & & \code{FALSE} & 2.28\\
    \hline
    \multirow{2}{*}{10000} & \multirow{2}{*}{608459} & \code{TRUE} & 266.00\\
    & & \code{FALSE} & 2.94\\
    \hline
    \multirow{2}{*}{17400} & \multirow{2}{*}{1074398} & \code{TRUE} & 842.2\\
    & & \code{FALSE} & 4.77\\
    \hline
  \end{tabular}
  \caption{Median run times for the [DBR] procedure at $\lambda = 0.05$.}
\end{table}


\end{appendix}


\end{document}